\documentclass[12pt,preprint]{aastex}
\usepackage{lscape}

\shorttitle{Circumstellar Geometry of SN 2009ip}
\shortauthors{Levesque et al.}

\begin{document}
\title{The Peculiar Balmer Decrement of SN 2009ip: Constraints on Circumstellar Geometry}
\author{Emily M. Levesque$^{1,2}$, Guy S. Stringfellow$^1$, Adam G. Ginsburg$^{1,3}$, John Bally$^1$, \& Brian A. Keeney$^1$}

\begin{abstract}
We present optical and near-IR spectroscopic observations of the luminous blue variable SN 2009ip during its remarkable photometric evolution of 2012. The spectra sample three key points in the SN 2009ip lightcurve, corresponding to its initial brightening in August (2012-A) and its dramatic rebrightening in early October (2012-B). Based on line fluxes and velocities measured in our spectra, we find a surprisingly low $I($H$\alpha)/I($H$\beta)$ ratio ($\sim$ 1.3-1.4) in the 2012-B spectra. Such a ratio implies either a rare Case B recombination scenario where H$\alpha$, but not H$\beta$, is optically thick, or an extremely high density for the circumstellar material of $n_e > 10^{13}$ cm$^{-3}$. The H$\alpha$ line intensity yields a minimum radiating surface area of $\gtrsim$20,000 AU$^2$ in H$\alpha$ at the peak of SN 2009ip's photometric evolution. Combined with the nature of this object's spectral evolution in 2012, a high circumstellar density and large radiating surface area imply the presence of a thin disk geometry around the central star (and, consequently, a possible binary companion), suggesting that the observed 2012-B rebrightening of SN 2009ip can be attributed to the illumination of the disk's inner rim by fast-moving ejecta produced by the underlying events of 2012-A.
\end{abstract}

\footnotetext[1]{CASA, Department of Astrophysical and Planetary Sciences, University of Colorado 389-UCB, Boulder, CO 80309, USA}
\footnotetext[2]{Hubble Fellow; {\tt Emily.Levesque@colorado.edu}}
\footnotetext[3]{European Southern Observatory, ESO Headquarters, Karl-Schwarzschild-Strasse 2, 95748 Garching bei M\"{u}nchen, Germany; ESO Fellow}

\section{Introduction}
The object hereafter referred to as SN 2009ip was discovered on 2009 August 26 (UT dates are used throughout this paper) by Maza et al.\ (2009) and initially identified as a faint ($\sim$17.9 $\pm$ 0.3 mag) supernova candidate. Miller et al.\ (2009) soon noted that historical data revealed the presence of a transient at the SN 2009ip location with evidence of photometric variability stretching back several years, and suggested that this supernova candidate was in fact the outburst of a variable star. Berger et al.\ (2009) later confirmed this with optical spectroscopy, measuring relatively narrow (FWHM $\sim$550 km s$^{-1}$) Balmer emission features and a relatively low absolute magnitude of $R \sim - 13.7$ and thus identifying SN 2009ip as an outbursting luminous blue variable (LBV) associated with the Sb spiral NGC 7259 ($D = 24$ Mpc; $z = 0.005714$), located 43\farcs7 (5.1 kpc in projection) from the host nucleus. Li et al.\ (2009) reported that this outburst showed a great deal of variability, noting that the LBV had faded to 20.2 mag on 2009 September 11 and then rapidly rebrightened to 18.3 mag on 2009 September 23. Both Smith et al.\ (2010) and Foley et al.\ (2011) later analyzed SN 2009ip, identifying it as a very massive ($\gtrsim$60 $M_{\odot}$) star showing evidence of repeated eruptions typical of an LBV based on pre-outburst archival photometry. Smith et al.\ (2010) also note that the fading and rebrightening behavior seen in the SN 2009ip outburst is extreme but comparable to the behavior of other LBVs.

Spectroscopy of SN 2009ip from Smith et al.\ (2010) during this first observed eruption event corresponded to a full width at half maximum (FWHM) velocity of $\sim$550 km s$^{-1}$ in the Balmer emission lines. However, contemporaneous spectroscopy from Foley et al.\ (2011) measured FWHM velocities of 780-890 km s$^{-1}$ in the weeks following the explosion, with later spectra showing broad absorption features with a maximum expansion velocity (measured in the blue wing of the absorption component) of $\sim$4500-7000 km s$^{-1}$, larger than the velocities measured for any other LBV eruption except the 1843 eruption of $\eta$ Carinae (Smith 2008). Spectroscopy from this initial outburst by Pastorello et al.\ (2013) also indicate a maximum expansion velocity of 5000-6000 km s$^{-1}$ for the fast-moving ejected material and FWHM velocities of 700-800 km s$^{-1}$, increasing to 1100-1200 km s$^{-1}$ at later times as the object faded back to quiescence after rebrightening.
 
A year after the 2009 outburst, Drake et al.\ (2010) reported additional activity from SN 2009ip detected in the Siding Spring Survey (SSS) data. An outburst was detected on 2010 July 15, once again followed by a fading period beginning on 2010 September 11 and a second outburst on  2010 September 29, similar to the fading and rebrightening behavior timescale of the 2009 outburst. From a 2010 October 6 spectrum, Pastorello et al.\ (2013) measure an H$\alpha$ FWHM velocity of $\sim$1300 km s$^{-1}$ during this outburst. Drake et al.\ (2010) also noted that these recent  frequent outbursts were similar to the pre-core-collapse outburst of SN 2006jc (e.g. Pastorello et al.\ 2007, Foley et al.\ 2007) and stated that such behavior could be indicative of the imminent core-collapse and explosion of SN 2009ip.

Pastorello et al.\ (2013) present photometric evidence of an additional eruptive phase in mid-2011, extending from May to October and marked by similar luminosity peaks and declines to the 2009 and 2010 outbursts. They obtained two spectra during this period, on 2011 September 2 and September 24. While the first is characterized as a dormant phase of SN 2009ip, with a FWHM velocity of $\sim$940 km s$^{-1}$ for H$\alpha$, the second spectrum shows narrower profiles (790 km s$^{-1}$ for H$\alpha$) and the emergence of strong P Cygni absorption in the Balmer lines. The blue absorption wings of these profiles indicate a maximum velocity for the fast-moving material of $\sim$12500 km s$^{-1}$, the highest outflow velocity ever measured for an LBV eruption.

2012 July 24 marked the discovery of the most recent outburst of SN 2009ip since its 2009 discovery (Drake et al.\ 2012). Spectroscopy of this new outburst (hereafter referred to as 2012-A) from as early as 2012 August 8 once again show strong broad absorption features, with the blue absorption wings of the Balmer lines extending up to 14000 km s$^{-1}$ and Lorentzian emission profiles with FWHM velocities of $\sim$1380 km s$^{-1}$ (Pastorello et al.\ 2013). Additional observations from 2012 August 26 (Vinko et al.\ 2012), and from 2012 September 16-23 (Smith \& Mauerhan 2012a, Mauerhan et al.\ 2013), also showed Balmer lines with very broad P Cygni absorption profiles corresponding to maximum velocities of $\sim$10000 km s$^{-1}$  and $\sim$13000 km s$^{-1}$, respectively. Foley et al.\ (2012) obtained high-resolution spectroscopy of the outburst from 2012 August 26 and noted prominent Balmer emission features with a somewhat lower FWHM velocity of 640 km s$^{-1}$, but did not include discussion of any P Cygni absorption components. While these observations were described as characteristic of the spectra and velocities associated with Type IIn supernovae (Smith \& Mauerhan 2012a, Mauerhan et al.\ 2013), contemporaneous photometry initially showed no evidence of any increase in brightness, suggesting luminosities too low to be associated with normal core-collapse SNe (Margutti et al.\ 2012, Martin et al.\ 2012). Indeed, photometric monitoring actually showed SN 2009ip fading in mid-September by $\sim$1 mag (Martin et al.\ 2012, Mauerhan et al.\ 2013, Pastorello et al.\ 2013, Prieto et al.\ 2013), similar to the post-outburst fading behavior observed in 2009 and 2010.

However, SN 2009ip began to brighten extremely rapidly $\sim$7 days later, on 2012 September 23.60 (Brimacombe 2012). This brightening (hereafter 2012-B) continued over a period of two weeks. It spanned $\sim$3-4 mag, reached an apparent peak or plateau on 2012 October 7, and showed an evolution and maximum luminosity consistent with other Type IIn supernovae (see Prieto et al.\ 2013 for detailed discussion). During this same time, additional spectroscopy of SN 2009ip showed that the previously-observed P Cygni features had since disappeared (Smith \& Mauerhan 2012b, Gall et al.\ 2012, Bohlsen 2012, Pastorello et al.\ 2013). However, spectroscopy from 2012 October 15 and 2012 October 22 suggested a re-emergence of P Cygni profiles with comparable velocities of 11200-13100 km s$^{-1}$ (Jha et al.\ 2012, Childress et al.\ 2012).

A number of different explanations for the 2012-A outburst of SN 2009ip have been presented. Mauerhan et al.\ (2013) and Prieto et al.\ (2013) both suggest that the 2012-A event is the signature of the initial stages of SN 2009ip undergoing core-collapse as a low-luminosity supernova, similar to the faint class of Type II-P's (Pastorello et al.\ 2007). They cite the extreme velocities measured in their spectra as implausible for any non-terminal outburst.  Alternatively, Pastorello et al.\ (2013) cite the broad P Cygni profiles and $\sim$12500 km s$^{-1}$ measured in their 2011 spectra as evidence that LBVs can indeed produce high-velocity blastwaves during non-terminal eruptions. They instead propose that 2012-A is the product of a pulsational pair-instability eruptive event, rather than the core-collapse death of SN 2009ip, stating that the low luminosity of the event is also not consistent with the SN death of such a massive LBV. Fraser et al.\ (2013) also support this interpretation; they suggest that their observations of SN 2009ip in December 2012 (which strongly resembled the object's appearance in late 2009) and models of the CSM ejecta are both consistent with the pulsation pair-instability mechanism rather than a core-collapse event.

However, most studies of SN 2009ip share a common interpretation for the 2012-B rebrightening and associated spectroscopic variations. This appears to be the product of the fast-moving 2012-A ejecta - whether produced by a supernova or a non-terminal mass eruption - catching up to and interacting with slower circumstellar medium (CSM) material ejected by past outbursts. The timescale separating 2012-A and 2012-B agrees with this model based on measured velocities (e.g. Mauerhan et al.\ 2013, Prieto et al.\ 2013, Margutti et al.\ 2013). Determining the true nature of SN 2009ip and its 2012 behavior - particularly whether 2012-A represented a terminal event - is only possible through continued long-term monitoring and careful analyses of multi-epoch datasets. However, observations of SN 2009ip during its 2012 evolution can still be used to characterize the nature of this unusual object's CSM.

Here we present our own optical and near-IR spectra of SN 2009ip, obtained at three distinct points in its observed photometric evolution: 2012 August 30 (during the 2012-A outburst), 2012 October 2 (during the 2012-B rapid rebrightening), and 2012 October 9 (after the 2012-B peak/plateau). We detail our acquisition and reduction of the data (\S 2), and present a series of analyses examining the line profiles and velocities measured in our spectra and the resulting implications for the CSM and effective radiating surface of SN 2009ip during its 2012 evolution (\S 3). Finally, we consider the potential conclusions that can be drawn from this work and the progress that can be made towards understanding the geometry and underlying nature of this remarkable object (\S 4).

\section{Observations and Reductions}
We obtained a series of optical and near-infrared spectroscopic observations of SN 2009ip at Apache Point Observatory (APO) during its 2012 period of activity. The dates of our spectroscopic observations are illustrated in Figure 1, along with R- and K-band lightcurves from the literature. Our first observations, in late August, coincide with the 2012-A outburst. Later observations in early October correspond to the 2012-B rebrightening and peak/plateau phases.

\subsection{Optical Spectroscopy}
Our optical observations were obtained using the medium dispersion ($R \sim 5000$) Dual Imaging Spectrograph (DIS) on the APO 3.5m telescope. DIS has separate blue and red cameras with the light split between the two using a dichroic with transition wavelength near 5350\AA. The B400 grating was used with the blue camera providing a linear dispersion of  $\sim$1.85\AA/pixel. The R300 grating used with the red camera extends from the dichroic cutoff to around 1 $\mu$m at $\sim$2.26\AA/pixel dispersion. All observations were taken with the $1\farcs5$ slit.

Our first optical spectrum was acquired on 2012 August 30. On this date, SN2009ip was observed during bright Moon near minimum airmass from APO ($\sim$2.1). We obtained three sequential 1000s exposures under photometric conditions but with variable seeing of  $\sim 2 \arcsec - 3\arcsec$. The flux standard BD+28 4211 (Oke 1990) was observed immediately afterwards at $\sim$1.1 airmass. The APO 3.5m telescope is an alt-az design, and has no provision to track the slit at the parallactic angle. Under the marginal conditions experienced on August 30, when the LBV was around $\sim$17.5 mag, the slit was left in the E-W orientation and the bright galaxy to the West was used to help facilitate positional guiding during the bright Moon phase. This inevitably resulted in some loss of light below $\lesssim$5200\AA, although the red portion of this spectrum can still be considered robust.

We next observed SN 2009ip on 2012 October 2, after the LBV had undergone an apparent SN-like explosion. These observations fill an important gap in the published spectroscopic series for this object (see Mauerhan et al.\ 2013, Pastorello et al.\ 2013, Margutti et al.\ 2013), sampling SN 2009ip during a period of rapid change in its optical and infrared lightcurves. Seeing during the DIS exposures (3x240s) observed at an airmass of $\sim$2.4 was $\sim 2.3 \arcsec$; flux standard BD+25 4655 (Oke et al.\ 1990) was observed at an airmass of $\sim$1.1. Additional DIS spectra were obtained a week later on 2012 October 9, precisely at the brightness peak of the $V$, $R$, and $K$ lightcurves (see Figure 1), with SN 2009ip at an airmass of $\sim$2.1 and flux standard BD+33 2642 (Oke et al.\ 1990) at an airmass of $\sim$1.4. Seeing during these exposures (3x300s) was $\sim 1.\arcsec3 $ at 2.11 airmass. Both observations were conducted under photometric conditions. In addition, since SN 2009ip was substantially brighter ($\sim$3 mag) during these observations and guiding was no longer an issue, the slit was rotated to and maintained at $90\degr$ to the horizon during the observations, approximating the parallactic angle. Thus, the flux calibration for the blue spectra on these dates should be more accurate than for August 30.

Extraction and reduction of the DIS spectra were performed using IRAF tasks, and included bias subtraction, flat fielding using quartz lamps, and wavelength calibration utilizing He, Ne, and Ar lamps that were supplemented by sky lines at the blue end. Flux calibrations for each SN 2009ip observation were performed using standard IDL routines and the observed spectrophotometric standard stars for each corresponding night. The spectra were also corrected for foreground reddening (E($B-V$) = 0.019; Schlegel et al.\ 1998). When merging the red and blue spectra, preference in duplicated spectral regions was given to the red spectrum because of its higher S/N and overall better flux calibration near the dichroic. No adjustment to the flux of either spectrum was made. The reduced spectra provide wavelength coverage extending from $\sim$3480\AA\ to the dichroic cutoff ($\sim$5350\AA), and from the dichroic cutoff to $\sim$9120\AA; these limits are not dictated by the sensitivity of the cameras, but instead by the low observed signal at the blue end and available data for performing the flux calibration using the observed standards at the red end.

\subsection{Near-IR Spectroscopy}
Near-infrared spectra were obtained on two dates at APO, 2012 August 31 and 2012 October 2. The medium-resolution cross-dispersed spectrograph TripleSpec (Wilson et al. 2004) was used, providing simultaneous spectral coverage from roughly 0.98-2.46 $\mu$m in 5 spectral orders. This yields a resolving  power R$\sim$3200 using a $1\farcs1 \times 43 \arcsec$ slit. Exposure times were 180s  for 2012 August 31 and 120s  for 2012 October 2, and were obtained by nodding along two positions on the slit, facilitating sky subtraction and correction of detector artifacts. Total on-source integration times were 2160s and 600s, respectively.

Sky subtraction, flat fielding, wavelength calibration using OH sky lines, and extraction of the spectra were achieved using  a modified version of the software package {\tt xspextool} (originally developed for use with SpeX on the NASA Infrared Telescope Facility; Cushing et al.\ 2004) redesigned for use with the specific instrument characteristics of TripleSpec. A $2\farcs5$ aperture extraction was used, with spatial background taken between $3 \arcsec - 6 \arcsec$ with a linear fit. The A0V stars HD 181801 and HD 201202 were used as flux and telluric standards on the respective nights. Telluric corrections and flux calibration were performed using the routine {\tt xtellcor} utilizing the method developed and described by Vacca et al.\ (2003).

\section{Analyses} 
\subsection{Identification and Fitting of the Emission Features}
Our optical and near-IR spectra are shown in Figure 2. Assuming the continuum can be characterized by a single (dominant) temperature for the emitting region, a blackbody fit can be attempted to our spectra. A 10,000 K blackbody fit for the $>$5400\AA\ regime with the strong lines masked out is shown for each spectrum in Figure 2. This provides a good fit to the 2012 October 9 spectrum above 5400\AA, though there is excess emission in the blue. Both the optical and near-infrared spectra from 2012 October 2 show good agreement with a 10,000 K blackbody continuum, although there appears to be some excess emission in the K-band that could indicate the initial stage of dust forming in the circumstellar environment.  Further observations are required to confirm the phenomenological explanation for either excess. The early 2012 August 30-31 spectra are not well represented by a 10,000 K blackbody (cooler fits also remain poor). However, this can be attributed to the loss of light below $\sim$5200\AA\ due to atmospheric dispersion effects and the non-parallactic-angle nature of the observations, which contributes to uncertainties in the blackbody fitting for these spectra.

Figure 3 illustrates the changes observed in the fluxes of the hydrogen Balmer lines as a function of time. Both the continuum flux and the emission line fluxes increase dramatically from the 2012-A period to the start of the 2012-B rebrightening, and show a further increase in flux between 2012 October 2 and 2012 October 9 as SN 2009ip reached its peak 2012-B brightness.

All three epochs of observations display strong emission lines of {\it only} the hydrogen Balmer series and helium; key line identifications are listed in Table 1. The integrated emission component of the line flux (F) is given in Table 1, along with the continuum-subtracted peak line flux (F$_{\rm peak}$). Fitting of line profiles in our data was done using the {\tt splot} routine in IRAF. These fits were performed for both normalized spectra (to better identify the continuum on either side of the lines) and unnormalized spectra for comparison. The main distinction lies in whether the defined continuum takes into account the uncertain ``absorption" components on the blue side of some lines. We have fit the continuum over a longer baseline, focusing on fitting the broad and narrow emission line components only. Line fitting was performed using a single component first, either a Gaussian or a Lorentzian profile. When a single-component fit was inadequate, then a two-component fit comprised of a Gaussian (narrow component) and a Lorentzian (broad component) profile was performed simultaneously (it is important to note that, due to the uncertain nature of the SN 2009ip CSM, these profiles merely represent the best geometric fit and cannot be used for a detailed interpretation of the underlying physical processes). Full-width zero-intensity velocities ($v_{\rm FWZI}$) attributed to the broad emission component are also listed in Table 1. Evidence for the broad emission component exists for all dates observed, including the 2012 August 30 spectrum coinciding with the 2012-A brightening, and occur in both the hydrogen Balmer and He I lines. The width of the broad component does vary between the epochs observed, apparently increasing during the 2012-B event. The measured FWHM velocities of the narrow emission line components are also listed in Table 1 and typically range from 400-800 km/s at all epochs. 

Fitting was also performed for the IR emission lines, which are quite narrow compared to the optical lines. The resulting velocities, both FWZI and FWHM, were substantially smaller than those observed in the optical. The FWHM line widths of only $\sim$200 km/s are typical of the winds produced by LBVs. 

Figure 4 displays the temporal evolution of the H$\alpha$, H$\beta$, and He I 5876\AA\ emission lines. A broad component of intermediate line widths ranging up to several thousand km s$^{-1}$ is present in the H$\alpha$  profile even during the 2012-A period, but is much less pronounced or even absent in the H$\beta$ profile for this epoch. The broad component feature becomes progressively more pronounced during the spectral evolution that coincides with the 2012-B rebrightening. This is accompanied by a significant and progressive increase in F$_{\rm peak}$ at each subsequent epoch. There is a significant increase in the FWZI between the 2012-A and 2012-B events, but little change between the 2012 October 2 and October 9 profiles beyond the substantial rise in F$_{\rm peak}$. The FWHM is marked in each panel. Perhaps what is most striking in the evolution of the H$\alpha$ and H$\beta$ profiles is the lack of any substantial Doppler shift between the various epochs, and the consistently symmetric line profiles.

The situation is somewhat different for the evolution of the He I 5876\AA\ line. It is not present in our first epoch of observations during the 2012-A period, but has already become strong before peak light for 2012 October 2, and continues to increase substantially in strength within the following week. This line is asymmetric, depressed on the blue side with a skewed broad red component. It is possible this could be due in part to blending with an unresolved nearby line to the red. However, there is clearly a very broad component by October 9 that extends to a FWZI of  $\sim$7000 km/s. The degree to which a P Cygni absorption component may contribute to the blue depression is unclear. The progressive increase in the strength of the He I 5876\AA\ line indicates that the temperature at the velocities where this line is forming has increased substantially between each subsequent epoch. This is also evident in the plots of the H$\alpha$ and H$\beta$ profiles, in which the lines of He I 4922\AA, 7065\AA, and 7282\AA, all of which are weak to nonexistent in the 2012 August 30 spectrum, have continuously increased in strength at later epochs.

\subsection{Implications of the Observed Balmer Decrement}
The relative line strengths of the H$\alpha$ and H$\beta$ emission lines provide constraints on the physical conditions in the emitting region. The continuum-subtracted spectrum provides a measure of Balmer line intensities and the Balmer decrement $D = I(H\alpha) / I(H\beta)$. In the absence of reddening, Case B recombination (the usual situation in photo-ionized nebulae when Ly$\alpha$ is optically thick but H$\alpha$ is optically thin) results in a ratio of D $\approx$ 3.0;  D depends weakly on temperature and varies from 3.3 at 2500 K, to 3.05 at 5,000 K, 2.87 at 10,000 K, and 2.76 at 20,000 K (Osterbrock \& Ferland 2005). These values apply to HII regions with densities up to $n_e \sim 10^6$ cm$^{-3}$. Reddening can only increase the Balmer decrement since H$\beta$ is attenuated more than H$\alpha$ by normal interstellar dust, and the foreground reddening in the direction of SN 2009ip is low, with E($B-V$) = 0.019 (Schlegel et al.\ 1998).

Drake \& Ulrich (1980) calculated the expected Balmer decrements for densities between $10^8 < n_e < 10^{15}$ cm$^{-3}$, a range of electron temperatures $5 \times 10^3 < T_e(K) < 4 \times 10^4$, and a large range in radiation fields and hydrogen line optical depths. Their results show that at densities between $10^8 < n_e < 10^{13}$ cm$^{-3}$, the Balmer decrement increases to values peaking between 10 to 15 around a density $n_e \approx 10^{10}$ to $10^{12}$ cm$^{-3}$ depending on $T_e$. The decrement then starts to thermalize due to collisions and decreases to near unity at $n_e \gtrsim 10^{13}$ cm$^{-3}$.  A simple estimate shows that the Balmer line intensities should approach a thermalization around this density. For a mostly ionized plasma with a temperature of $10^4$ K, the critical density required to thermalize the Balmer lines is roughly $n_{crit} \sim A_{ul} / \sigma_3 V_e \sim 10^{14}$  cm$^{-3}$ where $A_{ul} = 4.41 \times 10^7$ s$^{-1}$ is the Einstein A coefficient for the n = 3$-$2 transition in hydrogen, $\sigma_3 \sim 6.4 \times 10^{-15}$ is the Bohr cross-section of hydrogen in its n = 3 state, and $V_e \sim 400$ km s$^{-1}$ is the velocity of electrons, the most likely collision partner at $10^4$ K.

In SN 2009ip, the $I(H\alpha) / I(H\beta)$ ratio evolved from D $\sim$ 3 $\pm$ 0.2 on  2012 August 30 to D = 1.4 $\pm$ 0.1 on 2012 October 2 and D = 1.3 $\pm$0.1 on 2012 October 9. This calculation of the Balmer decrement is based on the ratio of the measured {\it peak} flux ($F_{\rm peak}$) for H$\alpha$ and H$\beta$ rather that the lines' integrated fluxes. While the geometric and velocity structure of the emitting region of SN 2009ip is unknown, it is likely that the lines' geometry is dictated by the extremely high velocities present in the circumstellar environment and, as a result, is likely produced by spatially-separated emission regions. As a result, the use of $F_{\rm peak}$ in determining D allows for an effective and consistent comparison between the low-velocity regions sampled by the line centers. However, adopting the integrated fluxes of the lines still produces extremely low values for the Balmer decrement: D= 1.7 $\pm$ 0.1 on 2012 October 2 and D = 1.6 $\pm$ 0.1 on 2012 October 9.

Following Drake \& Ulrich (1980), such a low Balmer decrement suggests that the line ratios are starting to approach LTE conditions, implying that the bulk of the emission is produced in a very high density plasma with $n_e > 10^{13}$ cm$^{-3}$. Alternatively, it is possible that a high optical depth in $H\alpha$ accompanied by a low optical depth in H$\beta$ could permit a relative increase in the brightness of H$\beta$ despite a decrease in the H$\beta$ emissivity (see Drake \& Ulrich 1980, Xu et al.\ 1992). This interpretation of a high optical depth for H$\alpha$ is supported by the models of Fraser et al.\ (2013); indeed, they find that their measured value of the H$\gamma$/H$\beta$ ratio (from a spectrum taken on 2012 October 8) is inconsistent with what is expected from an extreme high-density environment.

This is not the only supernova to display evidence of a shallow Balmer decrement; Chugai et al.\ (2004) report a Balmer decrement of D $\sim$ 1.1-1.2 in post-explosion spectra of SN 1994W taken at 21 and 31 days after the peak, with the decrement increasing slightly to D $\sim$ 1.5 by 49 days after peak. They note that while the narrow components of the emission features yield a nearly normal Balmer decrement, the broad components produce a flat or even inverse Balmer decrement, and speculate that this may indicate radiative transitions or collisional thermalization (see also Dessart et al.\ 2009).

\subsection{Calculating the Radiating Surface Area and Effective Radius}
The observed Balmer line fluxes constrain the radiating surface area. The peak H$\alpha$ flux (maximum intensity of the line at its peak, as opposed to the integrated area) from the 2012 October 9 spectrum of SN 2009ip is I(H$\alpha$) = $3.05 \times 10^{-14}$ ergs s$^{-1}$ cm$^{-2}$ \AA$^{-1}$, corresponding to a monochromatic H$\alpha$ luminosity of L(H$\alpha$) = $ 2 \times 10^{39}$ ergs s$^{-1}$ \AA$^{-1}$ at a distance of 24 Mpc. Assuming that this luminosity is produced by a disk of radius $R(H\alpha)$ at a blackbody temperature of $T = 10^4$ K, the {\it minimum} implied radius of the emitting region is given by $R(H\alpha) \gtrsim ( L(H\alpha) / 4 \pi ^2 B_{\nu}(T) \Delta \nu )^{0.5}$ $\gtrsim 1.2 \times 10^{15}$ cm, or approximately 80 AU. This in turn yields a minimum radiating surface area of $\sim$20000 AU$^2$.  Here, $\Delta \nu = 7 \times 10^{10}$ Hz is the bandwidth corresponding to 1\AA.

\section{Discussion}
Analyses of SN 2009ip and its 2012 have revealed the presence of extremely fast-moving ejecta produced during the 2012-A event. The origins of this fast-moving material are unclear, potentially produced either by the core-collapse death of the LBV (Mauerhan et al.\ 2013, Prieto et al.\ 2013) or a massive eruption such as a pulsational pair-instability event (Pastorello et al.\ 2013, Fraser et al.\ 2013). However, there is a consensus that the extraordinary brightening of 2012-B can be explained by the interaction of the fast-moving ejecta with much slower-moving ejecta from earlier LBV eruptions (e.g. Margutti et al.\ 2013). Here we focus on the interaction driving the 2012-B rebrightening, and consider two possible geometries:

{\bf Shell scenario:} The 2012-B rebrightening traces a rapidly expanding spherical shell powered by the fast-moving ejecta, or

{\bf Disk scenario:} The 2012-B rebrightening traces the inner edge of a dense circumstellar excretion disk illuminated by the fast moving ejecta.

In the shell scenario, earlier LBV mass loss events produce roughly spherical shells expanding at velocities on the order of half the H$\alpha$ FWHM velocity at the time of ejection. If formed during prior 2009 and 2010 eruptions, the expected velocities are around 500 to 1000 km s$^{-1}$ and reach radii of 5-10$\times10^{15}$ cm at the time of the 2012-B rebrightening. Fast ejecta coinciding with the 2012-A brightening ($\sim$13000 km s$^{-1}$ from Mauerhan et al.\ 2013), produced by either a core-collapse SN or an exceptionally powerful LBV eruption, would catch up with an older shell at this radius at the precise time scale that we observe, with the 2012-B rebrightening occurring $\sim$40 days after 2012-A. If the ejected masses are of order $\sim$1 M$_{\odot}$, the kinetic energy released by the impact of the fast ejecta on the older, slow ejecta could produce the radiated power observed in the 2012-B event. Dessart et al.\ (2009) proposed a similar scenario of interacting shells for SN 1994W, which displayed a shallow Balmer decrement similar to our spectra in early observations (see also Chugai et al.\ 2004).

However, the spectral behavior we see in SN 2009ip is inconsistent with this scenario. First, it is hard to understand the change in the Balmer decrement and the large implied density of the H$\alpha$ emitting region during the 2012-B rebrightening. Second, the impact of the fast SN or LBV ejecta is expected to be first seen as a highly blue-shifted component from the near (approaching) side of the slower spherical shell, with a light-crossing time delay of $\sim$2 days before showing a corresponding redshifted signature from the rear (receding) side (see, for example, the evolution of the SN 1987A H$\alpha$ spectrum from Kleiser et al.\ 2011). In contrast, our evolving line profiles in Figure 4 show that the peak of the emission is within 50 km s$^{-1}$ of the rest velocity of NGC 7259, SN 2009ip's presumed host, consistent with {\it no} Doppler shift of the strongest emission.

Our Balmer line profiles also differ from those observed in SN 1994W and examined in Dessart et al.\ (2009). They note the presence of P Cygni absorption profiles with shifts in the maximum velocities of the absorption components, and attribute this shift to the formation of more optically-thick lines in a region at a larger radius and velocity from the photosphere, concluding that the CSM of SN 1994W shows a velocity increase at greater distances in this shell scenario. There is no such evolution seen in the absorption components of our Balmer lines - indeed, our observed spectral lines seem to lack any clear P Cygni profiles at all (in contract with spectra of SN 2009ip taken at other epochs, e.g. Mauerhan et al.\ 2013, Pastorello et al.\ 2013, Fraser et al.\ 2013), let alone the narrow features typical of Type IIn supernovae and the geometry proposed for SN 1994W. This presents an additional challenge to the possibility of a shell scenario geometry for SN 2009ip; following the analyses in Dessart et al.\ (2009) our line geometry would imply a relatively compact line emission region in this scenario since it lacks any signature of the expected strong velocity gradient expected for a high-velocity expanding shell.

Alternately, in the disk scenario, the bulk of the emission arises from the radiating inner edge of a circumstellar disk surrounding the LBV. The eruptive behavior of SN 2009ip is quite similar to the observed behavior seen in $\eta$ Carinae, and long-term photometric and spectroscopic observations of SN 2009ip imply a semi-periodic nature on the order of a year. Combined, this suggests that SN 2009ip could potentially be modeled as a binary system (see also the binary merger model for SN 2009ip of Soker \& Kashi 2013). When the more massive member of a binary evolves off the main sequence and expands, mass transfer onto the companion or Roche lobe overflow can result in a massive circumstellar and circumbinary ``excretion" disk (a configuration in which matter flows outward into the disk, as opposed to the inward settling of matter in an accretion disk; Webbink 1976). Such disks and rings have been directly observed around many massive stars (e.g. Smith, Bally, \& Walawender 2007); examples include SN 1987A, Sher-25 near NGC 3603, and SBW1 (Smith et al.\ 2013).

In this scenario, the shocks resulting from the interaction of LBV eruptions impact the inner rim of a high-density excretion disk, possibly enhanced by ionizing radiation emitted by the LBV eruption or supernova explosion itself and creating the intense Balmer and helium emission (see Figure 5, top, for a schematic illustration of this geometry). Assuming an inner ring radius of order $10^{15}$ to $10^{17}$ cm (ranging from our measured minimum radiating surface radius for SN 2009ip to the typical ring radii seen in Sher-25, SN1987A, and SBW1), when either ionizing radiation or a blastwave hits the inner rim of this very high density disk the resulting emission would be dominated by gas with low radial velocities within 50 km s$^{-1}$ of the host, as observed for SN 2009ip. In the case of the impact of a dense shell from an LBV eruption, or possibly debris from a supernova, entrained material in the fast wind or ejecta flow would produce faint high-velocity emission. Entrained material located along the line of sight to bright continuum sources such as the remnant stellar photosphere, or hot plasma emission from the inner disk rim, would be seen in absorption, perhaps explaining the high-velocity P-Cygni profiles observed in SN 2009ip (Mauerhan et al.\ 2013).

Our observations of the SN 2009ip spectrum support this proposed geometry. The peak intensity of the H$\alpha$ line provides a constraint on the minimum surface area radiating this recombination line. Following our results in \S3.3,  the effective H$\alpha$ radiating surface area on 9 October 2012 in the peak of the H$\alpha$ profile was $A_{rad} \ge 4 \times 10^{30}$ cm$^2$ (20,000 AU$^2$).  If the disk inner edge has a radius $r$ and thickness $h$, these parameters have to satisfy the constraint $2\pi r h > A_{rad}$. The observed H$\alpha$ emission could arise from the inner ring edge if the ring has a radius larger than $\sim10^{15}$ cm, implying a thin ring. The high densities implied by our observed flat Balmer decrement in the spectra taken during the 2012-B rebrightening would therefore be a natural consequence of this circumstellar geometry. The timescale implied by this geometry also coincides perfectly with the delay required for illumination of the disk's inner ring by the 13000 km s$^{-1}$ 2012-A ejecta. Finally, at such high densities all forbidden line emission would be collisionally suppressed, as observed. Fraser et al.\ (2013) also conclude that the ``disk scenario" geometry is an effective means of explaining the appearance of the SN 2009ip spectra, although it should be noted that their interpretation is based on lower densities (10$^9$-10$^{10}$) determined from their own measurements of hydrogen line ratios.

The disk geometry scenario is also supported by other recent work on SN 2009ip. Fraser et al.\ (2013) also find that an extremely high velocity gradient is required in the circumstellar environment to explain the nature of the H$\alpha$ line, further supporting the interpretation that the mass loss processes that produced SN 2009ip's circumstellar geometry are the result of previous eruptive events rather than a steady stellar wind. Similarly, Margutti et al.\ (2013) find evidence for asymmetries in the explosion of SN 2009ip, noting the detection of multiple distinct ejecta velocities and spectral components that suggest an asymmetric distribution for the CSM material.

Our observations do {\it not} necessarily preclude the possibility of a shell scenario for the circumstellar geometry (see Figure 5, bottom). However, in order for such a geometry to agree with our observations the circumstellar material must fulfill precise conditions that are not commonly observed.  The H$\alpha$ optical depth must be greater than 1, while H$\beta$ must remain optically thin.  This situation still describes ``Case B" recombination, but yields a scenario where not all H$\alpha$ photons reach the observer, while all H$\beta$ photons do. At higher optical depth, the H$\beta$ line becomes trapped and H$\beta$ photons escape as a pair of Pa$\alpha$ and H$\alpha$ photons in the ``Case C" limit, which is not consistent with our observation of low $I($H$\alpha)/I($H$\beta)$. Within this limb-brightened shell geometry, it is easier to achieve a large radiating surface area than in the disk geometry. If we assume the same inner radius as above, $r\sim300$ AU, the limb-brightened region then only needs to be about 10AU thick in order to provide the observed luminosity.

While this work can constrain the geometry of SN 2009ip's circumstellar environment, the origin of the fast-moving ejecta illuminating the circumstellar material remains unclear. While either geometry is consistent with ejecta produced during the 2012-A event, distinguishing between material produced by a core-collapse SN or an extreme LBV eruption is not possible from current data. Future observations of SN 2009ip's spectral and photometric evolution will shed important new light on the true circumstellar geometry of this star and also ultimately clarify the nature of the 2012-A behavior of SN 2009ip. \\

We gratefully acknowledge useful discussions and contributions from Charles Danforth, Kevin France, Cynthia Froning, and John Stocke. This work was based on observations obtained with the Apache Point Observatory 3.5-meter telescope, which is owned and operated by the Astrophysical Research Consortium. We thank the support staff at Apache Point for their expertise and assistance during these observations. This research made use of {\tt pyspeckit}, an open-source spectroscopic toolkit for Python hosted at http://pyspeckit.bitbucket.org. EML is supported by NASA through Einstein Postdoctoral Fellowship grant number PF0-110075 awarded by the Chandra X-ray Center, which is operated by the Smithsonian Astrophysical Observatory for NASA under contract NAS8-03060, and through Hubble Fellowship grant  number HST-HF-51324.01-A from the Space Telescope Science Institute, which is operated by the Association of Universities for Research in Astronomy, Incorporated, under NASA contract NAS5-26555. GSS gratefully acknowledges support from NASA Grant  NNX13AF34G, and BAK gratefully acknowledges support from NASA grant NNX08AC14G and NSF grant AST1109117.

\begin{figure}
\epsscale{1}
\plotone{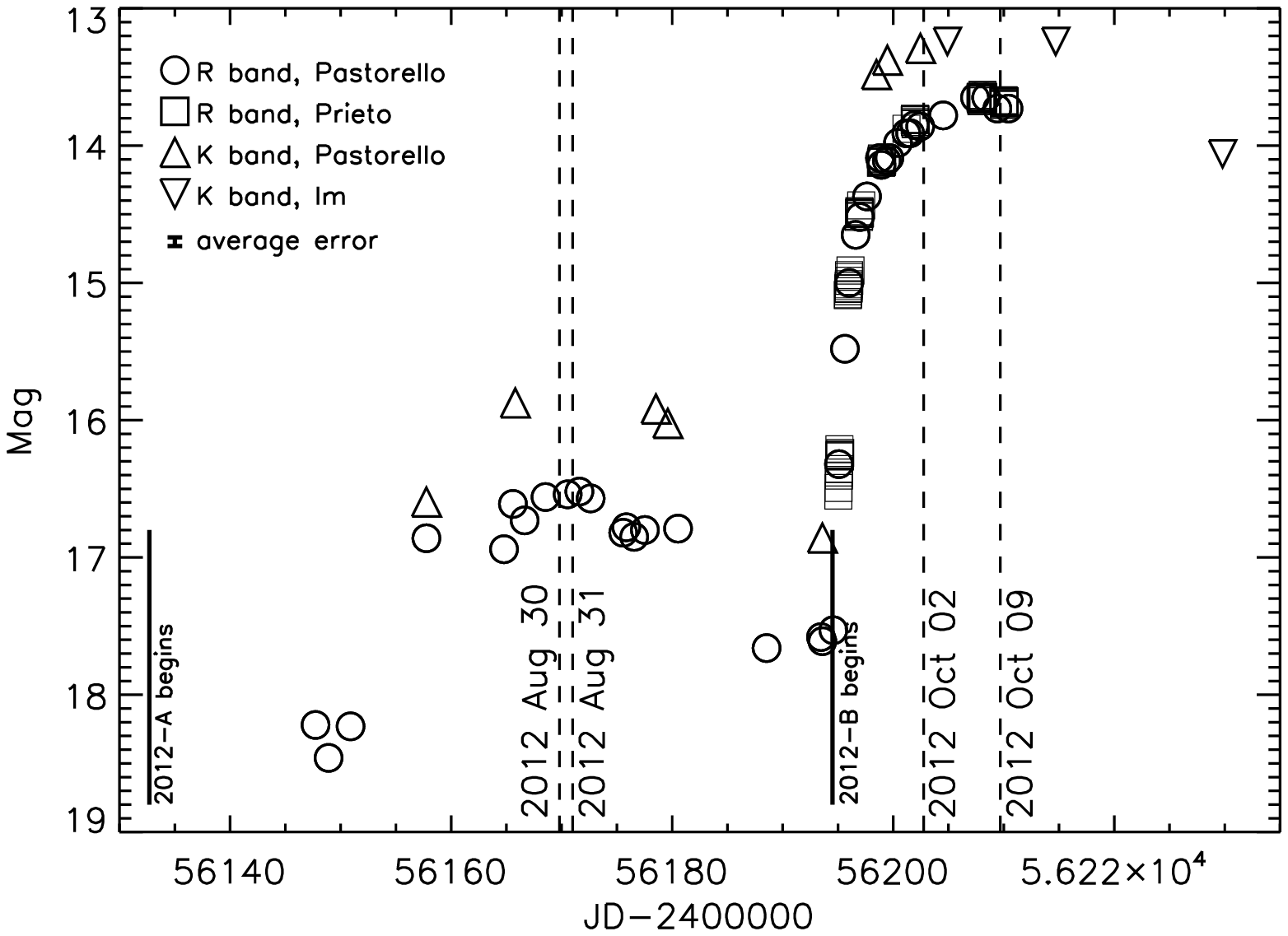}
\caption{R and K photometry of SN 2009ip during late 2012 from Pastorello et al. (2013), Prieto et al.\ (2013), and Im (2012). The dates of our APO spectroscopy observations are indicated as dashed lines, illustrating their temporal relation to the evolution of the SN 2009ip lightcurve. The earliest detection dates for the 2012-A outburst and 2012-B rebrightening are shown as solid lines. The dotted line shows the average rate of change in the K-band lightcurve from Im (2012)}
\end{figure}

\begin{figure}
\epsscale{1}
\plotone{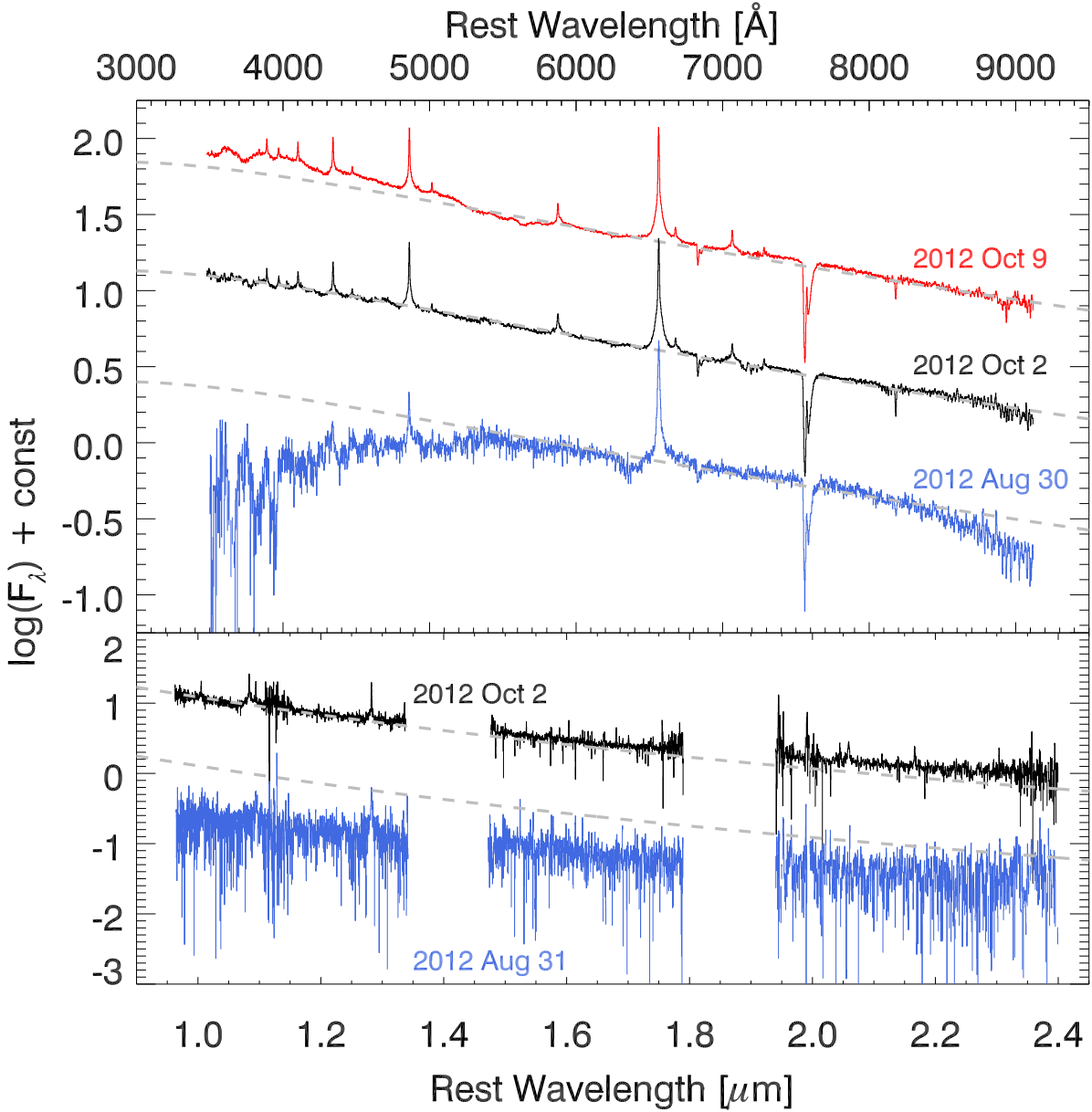}
\caption{Apache Point Observatory 3.5-m spectrophotometry of SN 2009ip taken with DIS (top) and TripleSpec (bottom), showing data from 2012 August 30-31 (blue), 2012 October 2 (black), and 2012 October 9 (red). The spectra have been corrected to rest-frame wavelengths. The dashed gray lines illustrate the agreement of each spectrum with a 10,000 K blackbody fit.}
\end{figure}

\begin{figure}
\epsscale{1}
\plotone{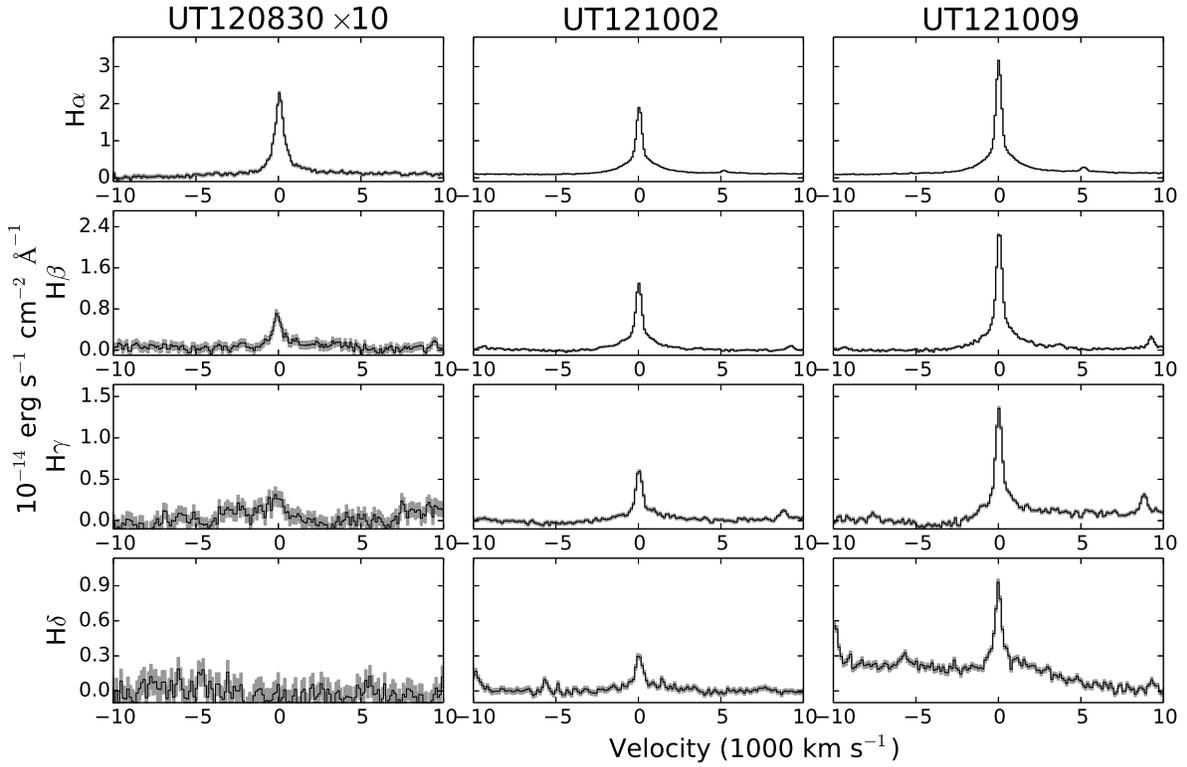}
\caption{Line profiles of the brightest hydrogen Balmer series emission features, showing changes in these line's fluxes and profiles (on a velocity scale) as a function of time. All spectra have had a linear continuum subtracted. The observations from 2012 August 30 are plotted with a multiplicative factor of ten to better illustrate the line profiles. Errors are shown as gray bars.} 
\end{figure}

\begin{figure}
\epsscale{0.55}
\plotone{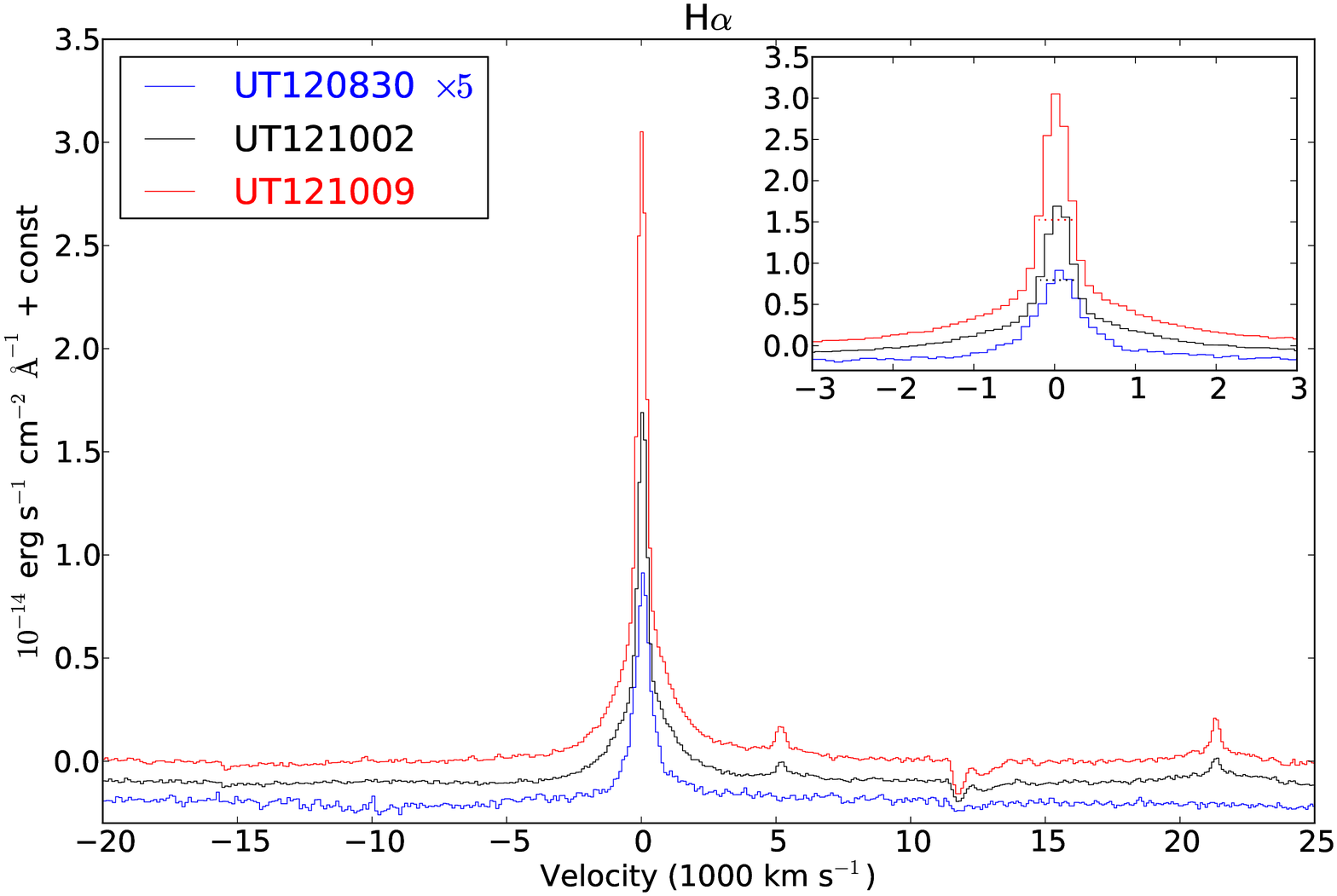} 
\plotone{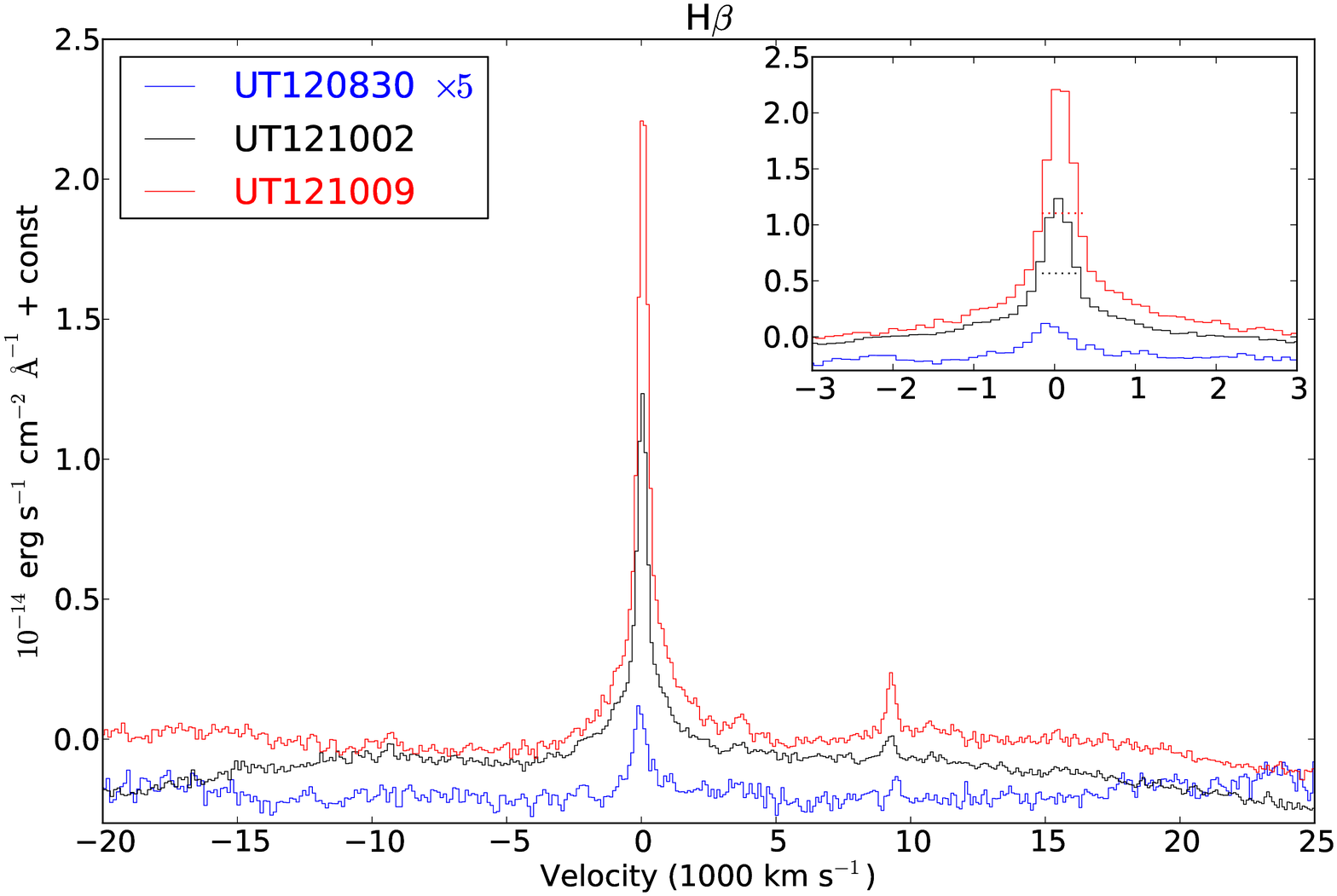}
\plotone{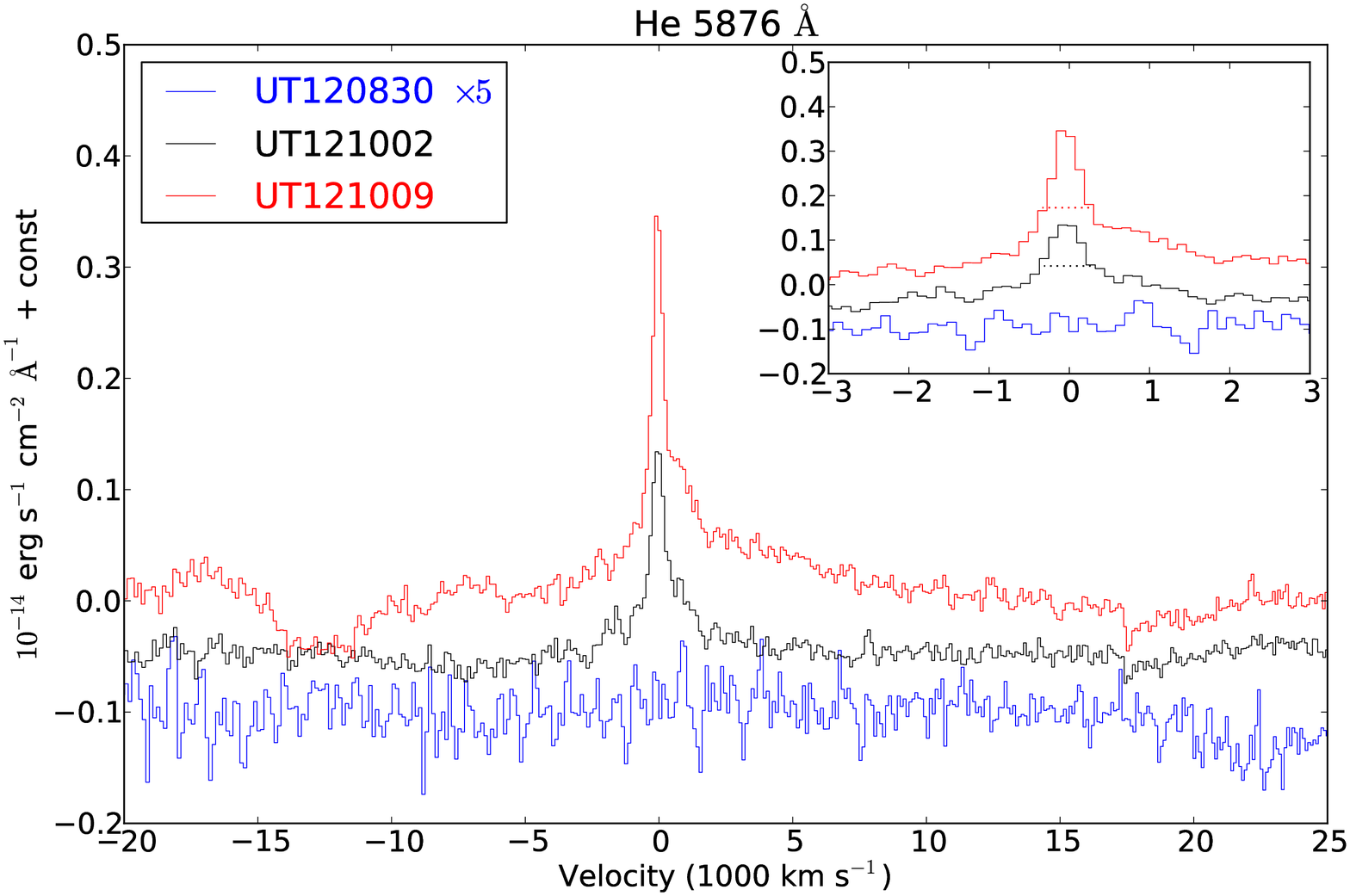}
\caption{Temporal evolution of the H$\alpha$ (top), H$\beta$ (middle), and He I 5876\AA\ (bottom) emission features in our spectra from 2012 August 30-31 (blue), 2012 October 2 (black), and 2012 October 9 (red). The evolution is shown for continuum subtracted line profiles, with the 2012 August 30 data scaled higher by a factor of 5 in order to highlight features at that fainter epoch. The smaller insert shows kinematics over a range of $\pm$ 3000 km s$^{-1}$, with the FWHM height \& width marked as a dotted line. All lines show an increase in strength with time. Note that in the H$\alpha$ and H$\beta$ panels, the weak He I 4922\AA, 7065\AA, and 7282\AA\ emission lines also continuously increase in strength at later epochs.}
\end{figure}

\begin{figure}
\epsscale{1}
\plotone{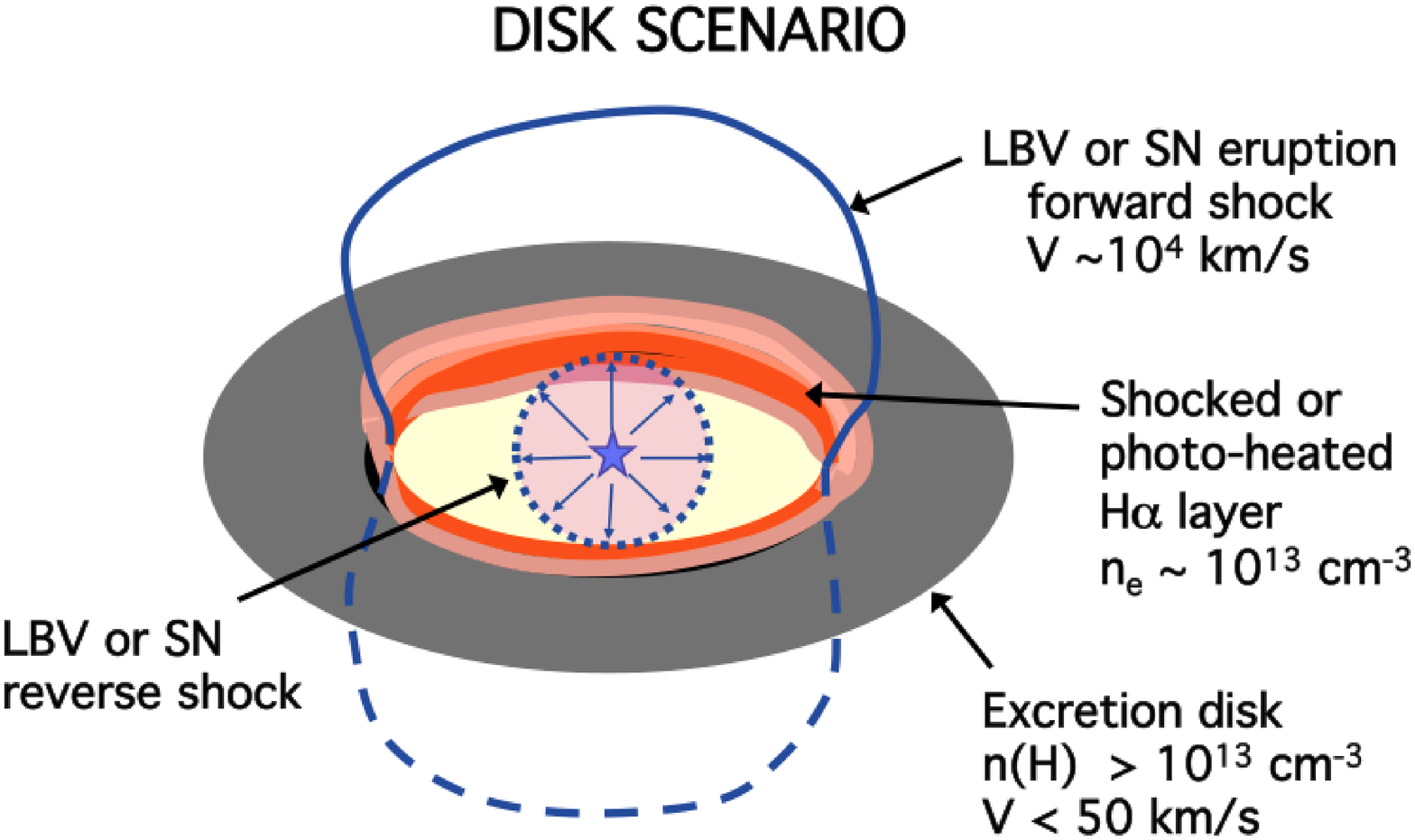}
\plotone{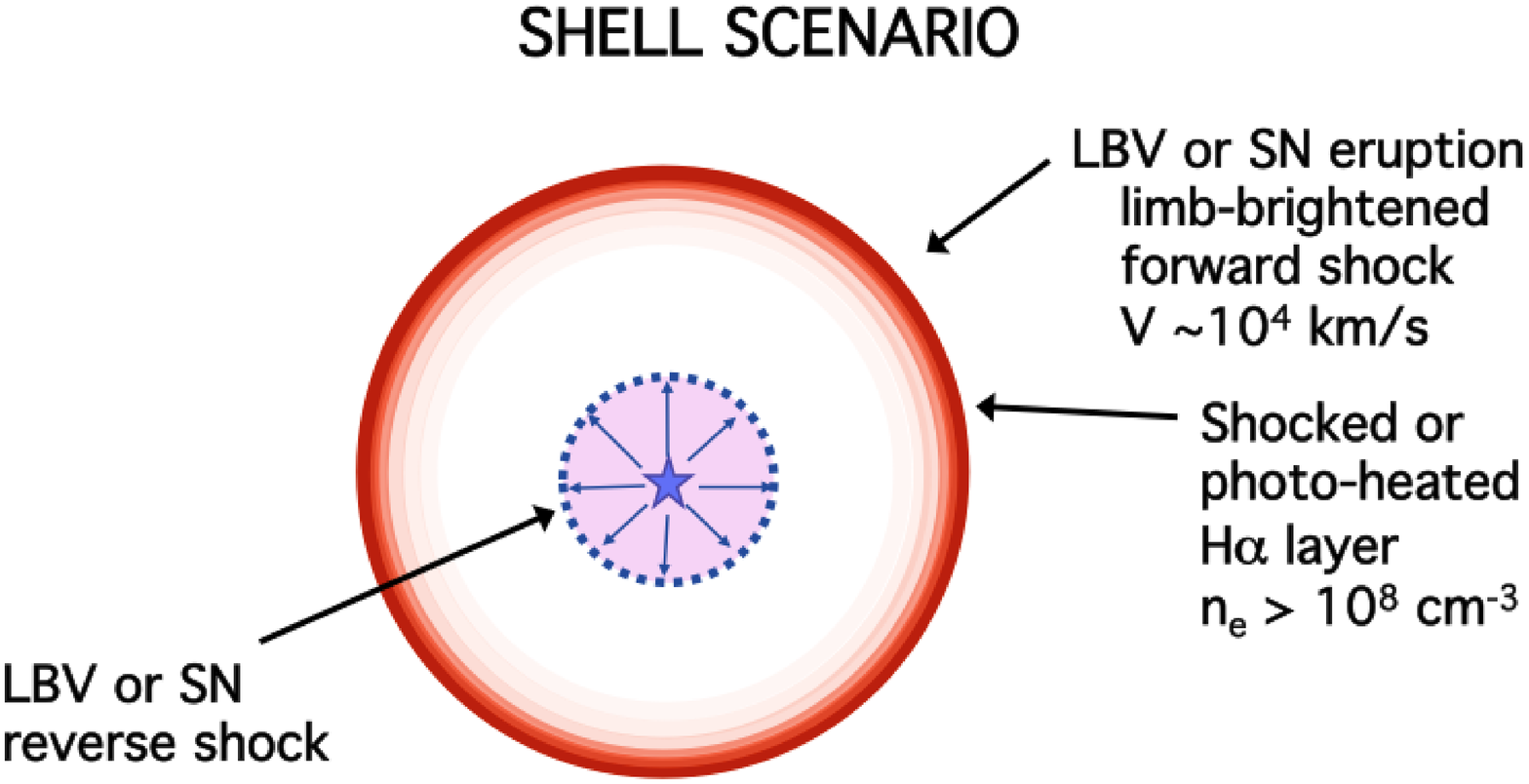}
\caption{Illustration of the two possible circumstellar geometry scenarios that can explain our observations of SN 2009ip; a high-density thin disk geometry (top) or a limb-brightened shell geometry where the H$\alpha$ optical depth is greater than 1 while H$\beta$ remains optically thin.}
\end{figure}

\begin{landscape}
\begin{deluxetable}{l c c c c c c c c c c c c c c c c c c}
\vspace{-100pt}
\tabletypesize{\scriptsize}
\tablewidth{0pc}
\tablenum{1}
\tablecolumns{19}
\tablecaption{\label{tab:gals} Key SN 2009ip Emission Features}
\tablehead{
\colhead{}
&\colhead{}
&\multicolumn{5}{c}{\bf 2012 Aug 30/31\tablenotemark{1}}
&\colhead{}
&\multicolumn{5}{c}{\bf 2012 Oct 2}
&\colhead{}
&\multicolumn{5}{c} {\bf 2012 Oct 9} \\ \cline{3-7} \cline{9-13} \cline{15-19}
\colhead{Species}
&\colhead{$\lambda_{\rm rest}$}
&\colhead{F\tablenotemark{2}}
&\colhead{F$_{\rm peak}$}
&\colhead{$v_{\rm \tiny FWHM}$\tablenotemark{3}}
&\colhead{$v_{\rm \tiny FWZI}$}
&\colhead{Fit\tablenotemark{4}}
&\colhead{}
&\colhead{F}
&\colhead{F$_{\rm peak}$}
&\colhead{$v_{\rm \tiny FWHM}$}
&\colhead{$v_{\rm \tiny FWZI}$}
&\colhead{Fit}
&\colhead{}
&\colhead{F}
&\colhead{F$_{\rm peak}$}
&\colhead{$v_{\rm \tiny FWHM}$}
&\colhead{$v_{\rm \tiny FWZI}$}
&\colhead{Fit} \\
\colhead{}
&\colhead{(\AA)}
&\colhead{ }
&\colhead{ }
&\colhead{(km s$^{-1}$)}
&\colhead{(km s$^{-1}$)}
&\colhead{}
&\colhead{}
&\colhead{ }
&\colhead{ }
&\colhead{(km s$^{-1}$)}
&\colhead{(km s$^{-1}$)}
&\colhead{}
&\colhead{}
&\colhead{ }
&\colhead{ }
&\colhead{(km s$^{-1}$)}
&\colhead{(km s$^{-1}$)}
&\colhead{}
}
\startdata
\hline
{\bf DIS} & & & & & & & & & & & & & & & & & & \\
\hline 
H$\delta$ &4101 	 &\nodata 	& \nodata & \nodata & \nodata & \nodata & & 3.9 & 0.26 & 610 & 3000 &L & & 6.3 & 0.72 & 480 & 1900 &L \\
H$\gamma$ &4341 	 & 0.30 & 0.028 & 730 & 800 &G & & 7.8 & 0.61 & 570 & 4500 &L & & 15.9 & 1.33 & 530 & 4400 &GL \\
H$\beta$ &4861 	 & 0.80 & 0.065 & 750 & 2000 &L & & 20.1 & 1.30 & 540 & 6000 &GL & & 29.3 & 2.2 & 500 & 6000 &GL \\
He I &5876 	 & \nodata & \nodata & \nodata & \nodata & \nodata & & 4.5 & 0.17 & 870 & 4800 &GL & & 5.1 & 0.30 & 620 & 3500 &GL \\
H$\alpha$ &6563 	 & 4.6 & 0.216 & 630 & 5000 &L & & 33.4 & 1.78 & 500 & 7000 &GL & & 45.7 & 2.9 & 450 & 6000 &GL \\
He I &7065 	 & \nodata & \nodata & \nodata & \nodata & \nodata & & 0.69 & 0.070 & 360 & 1200 &G & & 2.2 & 0.19 & 400 & 1200 &GL \\
\hline
{\bf TSpec} & & & & & & & & & & & & & & & & & & \\
\hline
Pa$\gamma$ & 10938 	 & 0.84 & 0.11 & 200 & 400 &G & & 0.83 & 0.10 & 200 & 400 &G & &$-$ &$-$ &$-$ &$-$ &$-$ \\ 
Pa$\beta$ & 12818 	 & 1.4 & 0.13 & 230 & 600 &G & & 1.6 & 0.13 & 230 & 700 &G & &$-$ &$-$ &$-$ &$-$ &$-$ \\ 
Br$\gamma$ & 21655 	 & 0.16 & 0.014 & 170 & 300 &G & & 0.17 & 0.013 & 180 & 300 &G & &$-$ &$-$ &$-$ &$-$ &$-$ \\ 
\enddata	   
\tablenotetext{1}{DIS data from 2012 Aug 30; TripleSpec data from 2012 Aug 31.}
\tablenotetext{2}{Units of 10$^{-14}$ ergs s$^{-1}$ cm$^{-2}$ \AA$^{-1}$; fluxes have measurement errors of $\pm$10\%.} 
\tablenotetext{3}{$v_{FZWI}$ values have measurement errors of $\pm$10\%.}
\tablenotetext{4}{Fit types; G: Gaussian, L: Lorentzian, GL: Gaussian+Lorentzian}
\end{deluxetable}
\end{landscape}

\end{document}